\documentclass[useAMS,usenatbib,onecolumn]{mn2e}
\usepackage{amsmath,bm}
\usepackage{dcolumn}
\usepackage[utf8]{inputenc}
\usepackage{graphics,graphicx}
\usepackage{float}

\usepackage{color}

\newcommand{\p}{^\prime}
\newcommand{\pp}{^{\prime\prime}}

\title[PH$_3$ probes of a variable $\mu$]{Anomalous phosphine sensitivity coefficients as probes for a possible variation of the proton-to-electron mass ratio}

\date{\today}

\author[Owens, Yurchenko and \v{S}pirko]
{A. Owens$^{1,2,3}$\thanks{The corresponding author: alec.owens@cfel.de}, S. N. Yurchenko$^{3}$ and V. \v{S}pirko$^{4,5}$\thanks{The corresponding author: spirko@marge.uochb.cas.cz} \\ \\
$^1$ The Hamburg Center for Ultrafast Imaging, Universit\"{a}t Hamburg, Luruper Chaussee 149, 22761 Hamburg, Germany\\
$^2$ Center for Free-Electron Laser Science (CFEL), Deutsches Elektronen-Synchrotron DESY, Notkestrasse 85, 22607 Hamburg, Germany\\
$^3$ Department of Physics and Astronomy, University College London, Gower Street, WC1E 6BT London, United Kingdom\\
$^4$ Academy of Sciences of the Czech Republic, Institute of Organic Chemistry and Biochemistry,\\
Flemingovo n\'am.~2, 166 10 Prague 6, Czech Republic \\
$^5$ Department of Chemical Physics and Optics, Faculty of Mathematics and Physics, Charles University in Prague,\\
Ke Karlovu 3, CZ-12116 Prague 2, Czech Republic}

\date{Accepted XXXX. Received XXXX; in original form XXXX}

\pagerange{\pageref{firstpage}--\pageref{lastpage}} \pubyear{2017}

\begin{document}

\label{firstpage}

\maketitle

\begin{abstract}
A robust variational approach is used to investigate the sensitivity of the rotation-vibration spectrum of phosphine (PH$_3$) to a possible cosmological variation of the proton-to-electron mass ratio, $\mu$. Whilst the majority of computed sensitivity coefficients, $T$, involving the low-lying vibrational states acquire the expected values of $T\approx-1$ and $T\approx-1/2$ for rotational and ro-vibrational transitions, respectively, anomalous sensitivities are uncovered for the $A_1\!-\!A_2$ splittings in the $\nu_2/\nu_4$, $\nu_1/\nu_3$ and $2\nu_4^{\ell=0}/2\nu_4^{\ell=2}$ manifolds of PH$_3$. A pronounced Coriolis interaction between these states in conjunction with accidentally degenerate $A_1$ and $A_2$ energy levels produces a series of enhanced sensitivity coefficients. Phosphine is expected to occur in a number of different astrophysical environments and has potential for investigating a drifting constant. Furthermore, the displayed behaviour hints at a wider trend in molecules of $\bm{C}_{3\mathrm{v}}\mathrm{(M)}$ symmetry, thus demonstrating that the splittings induced by higher-order ro-vibrational interactions are well suited for probing $\mu$ in other symmetric top molecules in space, since these low-frequency transitions can be straightforwardly detected by radio telescopes.
\end{abstract}


\section{Introduction}

 Recently, the $J\!=\!2\!-\!1$ rotational transition of phosphine (PH$_3$) was detected in the carbon star envelope IRC +10216~\citep{Agundez}, thus confirming the presence of PH$_3$ in the outflows of evolved stars but more significantly outside of the solar system. The appearance of PH$_3$ has been predicted in numerous other astrophysical environments (see the discussion by \citet{15SoAlTe.PH3} and references therein), and because of prominent `irregularities' displayed by its rotation-vibration spectrum, it is a promising system for investigating the cosmological variability of the proton-to-electron mass ratio, $\mu=m_p/m_e$. Observing PH$_3$ outside of our Galaxy is no easy feat, however, nearby Galactic molecular clouds offer a means to constrain $\mu$ through the so-called chameleon scenario~\citep{Khoury:2004,Brax:2004} as evidenced by studies of ammonia~\citep{Levshakov:2010b,Levshakov:2010a} and methanol~\citep{Dapra2017}.

 At present, the most robust constraint on a temporal variation of $\mu$ was determined from methanol absorption spectra observed in the lensing galaxy PKS1830$-$211~\citep{Kanekar:2015}. The three measured transitions possessed sensitivity coefficients, $T$, ranging from $-7.4$ to $-1.0$ and resulted in a constraint of $\dot{\mu}/\mu< 2\times 10^{-17}\,$yr$^{-1}$ assuming a linear rate of change. This translates to no change in $\mu$ over the past $\approx 7.5$ billion years and is in agreement with the best laboratory constraint to date, which measured optical transitions in $^{171}$Yb$^+$ ions to derive $\dot{\mu}/\mu= (0.2\pm 1.1)\times 10^{-16}\,$yr$^{-1}$~\citep{Godun:2014} again assuming a linear rate of change.  Whilst the use of methanol has led to several astronomical constraints~\citep{Jansen:2011,Levshakov:2011,Bagdonaite:2013b,Bagdonaite:2013a,Thompson:2013,Kanekar:2015}, it is worthwhile identifying other molecular absorbers with notable sensitivities to expand the search for a drifting $\mu$.

\begin{figure}
%hspace*{-20mm}
\centering
\includegraphics{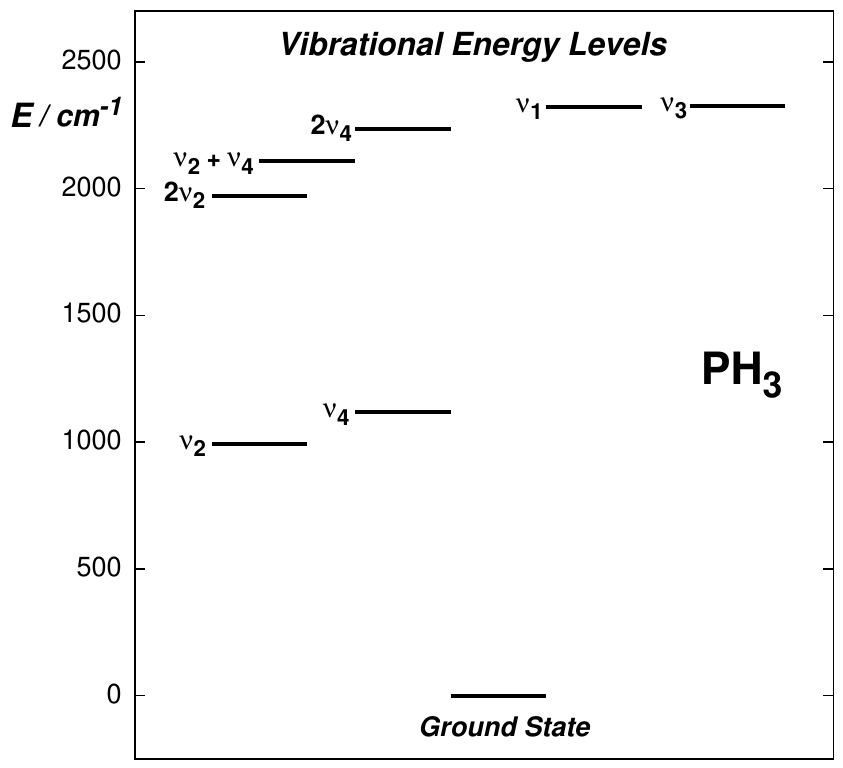}
\caption{\label{fig:one}The lowest vibrational energy levels of PH$_3$.}
\end{figure}

 Due to the small difference between its rotational constants $B$ and $C$, and also because of the strong $x\!-\!y$ Coriolis interaction between the coinciding $\nu_2/\nu_4$, $\nu_1/\nu_3$ and $2\nu_4^{\ell=0}/2\nu_4^{\ell=2}$ states (see Fig.~\ref{fig:one}), phosphine is a potential candidate system for probing $\mu$. Notably, the spectrum of PH$_3$, and presumably other molecules of $\bm{C}_{3\mathrm{v}}\mathrm{(M)}$ symmetry, is special due to the anomalous behaviour of the $A_1\!-\!A_2$ splittings~\citep{Ulenikov}. A large number of spectroscopic studies of PH$_3$ have been reported in the literature (see \citet{Mueller} and references therein) and highly accurate data is available for the majority of its states. Furthermore, a robust theoretical description of this molecule, which we utilize for this work, has been developed over the years~\citep{03YuCaJe.PH3,05YuThJe.PH3,06YuCaTh.PH3,
 OvThYu08a.PH3,OvThYu08.PH3,13SoYuTe.PH3,
 14SoHYu.PH3,15SoAlTe.PH3,16SoTeYu.PH3}, culminating in the construction of a comprehensive rotation-vibration line list applicable for elevated temperatures~\citep{15SoAlTe.PH3}.

 Model radiative transfer calculations of phosphine excitation in the envelope of IRC +10216~\citep{Agundez,Cernicharo} highlighted the importance of infrared pumping from the ground to the first excited vibrational states, helping explain the presence of strong emission bands in the observed spectra. We therefore find it useful to investigate the sensitivity of the ground, fundamental, and low-lying combination and overtone vibrational states of PH$_3$ (see Fig.~\ref{fig:one}) to a possible space-time variation of $\mu$ using a robust variational approach. The paper is structured as follows:  In Sec.~\ref{sec:methods} we describe the variational approach used to compute sensitivity coefficients. The results for the phosphine molecule are presented and discussed in Sec.~\ref{sec:results}. Concluding remarks are given in Sec.~\ref{sec:conc}.

\section{Variational Approach}
\label{sec:methods}

 The sensitivity coefficient $T_{u,l}$ between an upper and lower state with energy $E_u$ and $E_l$, respectively, is defined as
\begin{equation}
T_{u,l}=\frac{\mu}{E_u-E_l}\left(\frac{{\rm d}E_u}{{\rm d}\mu}-\frac{{\rm d}E_l}{{\rm d}\mu}\right),
\label{eq.T}
\end{equation}
and can be related to the induced frequency shift of a transition, or energy difference $E_u-E_l$ between two states, through the expression
\begin{equation}
\frac{\Delta\nu}{\nu_0}=T_{u,l}\frac{\Delta\mu}{\mu_0},
\label{eq.shift}
\end{equation}
where $\Delta\nu=\nu_{\mathrm{obs}}-\nu_0$ is the change in the frequency, and $\Delta\mu=\mu_{\mathrm{obs}}-\mu_0$ is the change in $\mu$, both with respect to their present day values $\nu_0$ and $\mu_0$. By assuming all baryonic matter can be treated equally~\citep{Dent:2007}, $\mu$ is proportional to the molecular mass. One can then perform a series of calculations with suitably scaled values for the masses of the P and H atoms and extract numerical values for the derivatives ${\rm d}E/{\rm d}\mu$ using central finite differences.

 Sensitivity coefficients for PH$_3$ have been computed with the same variational approach as was previously employed for ammonia~\citep{Owens:2015,Owens:NH3:PRA} and the hydronium cation~\citep{15OwYuPo.H3Op}. Calculations were carried out with the nuclear motion program \textsc{trove}~\citep{TROVE:2007,15YaYu.ADF,Symmetry:2017} and utilized the potential energy surface (PES), dipole moment surface (DMS), and computational setup of \citet{15SoAlTe.PH3}, which have all undergone rigorous testing and are known to be reliable. We refer the reader to \citet{15SoAlTe.PH3} for further details of the nuclear motion computations. All sensitivity coefficients, Eq.~\eqref{eq.T}, have been determined with calculated frequencies, $E_u-E_l$, as oppose to experimental values when available. This was done for consistency and to verify the trend in sensitivities displayed by PH$_3$, which we will discuss further in Sec.~\ref{sec:results}.

\section{Results and Discussion}
\label{sec:results}

\begin{figure}
\centering
\includegraphics{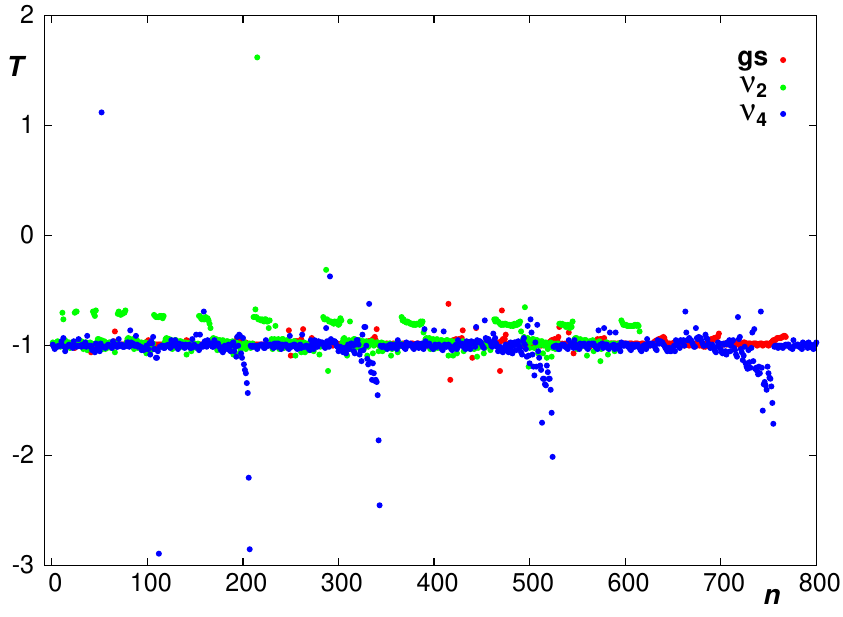}
\caption{\label{fig:two}Sensitivity coefficients $T$ for pure rotational transitions in the ground, $\nu_2$, and $\nu_4$ vibrational states of PH$_3$. Here $n$ is a running number which counts the number of transitions.}
\end{figure}

\begin{figure}
\centering
\includegraphics{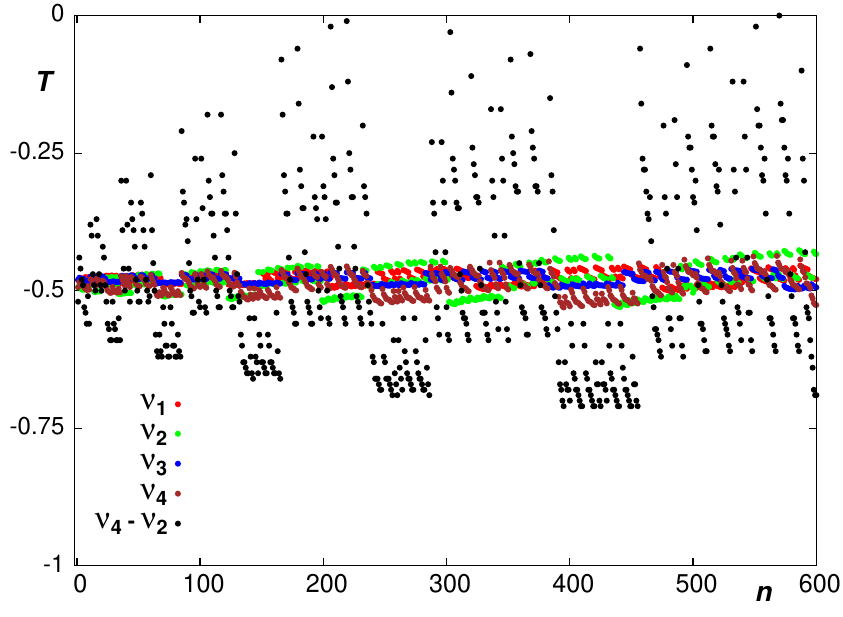}
\caption{\label{fig:three}Sensitivity coefficients $T$ for ro-vibrational transitions from the ground to the lowest vibrational states of PH$_3$. Here $n$ is a running number which counts the number of transitions.}
\end{figure}

\begin{figure}
\centering
\includegraphics{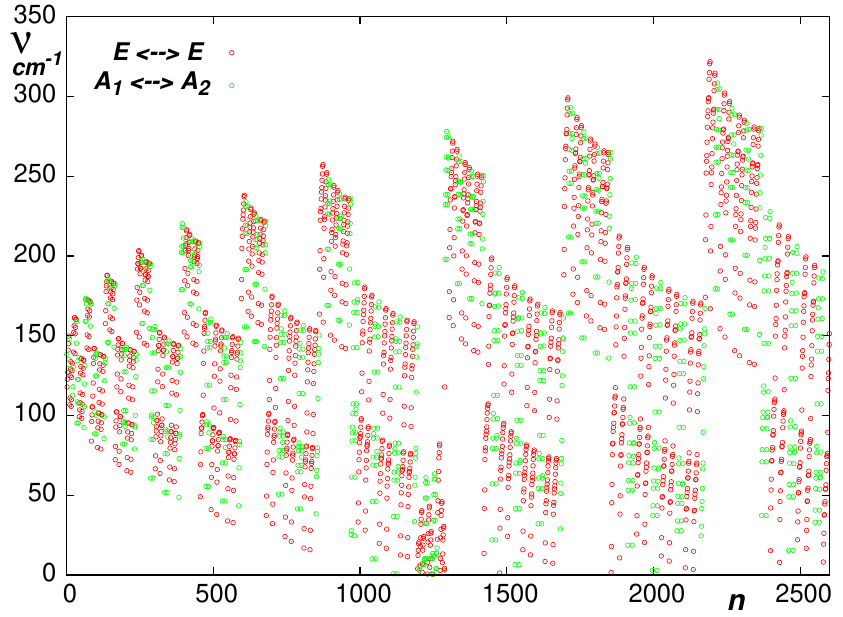}\\
\includegraphics{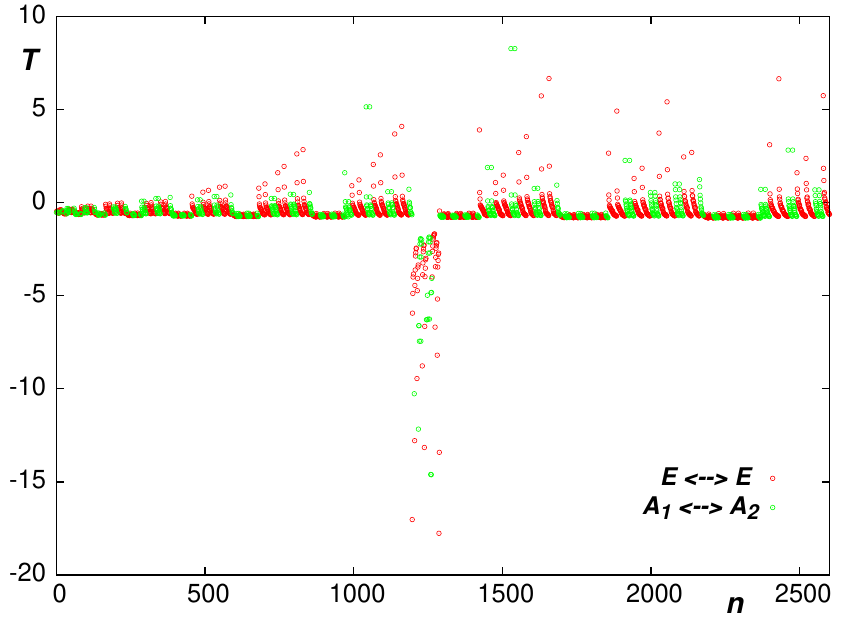}
\caption{\label{fig:four}The wavenumbers $\nu$ (in cm$^{-1}$) and sensitivity coefficients $T$ of the $\nu_4\leftarrow\nu_2$ ro-vibrational transitions of PH$_3$. Here $n$ is a running number which counts the number of transitions.}
\end{figure}

 In general, as shown in Table~\ref{tab:gs}, Fig.~\ref{fig:two} and Fig.~\ref{fig:three}, the majority of the calculated sensitivity coefficients for the low-lying vibrational states acquire the expected values of $T\approx-1$ and $T\approx-1/2$ for rotational and ro-vibrational transitions, respectively. Notably, this is the case for the $J\!=\!2\!-\!1$ and $J\!=\!1\!-\!0$ rotational transitions observed in the carbon star envelope IRC +10216~\citep{Agundez:2008,Agundez}. For a small fraction of the probed transitions the sensitivities deviate from the usual values. Accidental coincidences between ro-vibrational states can cause the magnitude of these `irregularities' to strongly increase with vibrational excitation, as illustrated in Fig.~\ref{fig:four}.

\begin{table}
\centering
\caption{\label{tab:gs}Calculated frequency $\nu_{\mathrm{calc}}$ (in MHz), frequency difference $\Delta_{\mathrm{c}-\mathrm{e}}$ (in MHz) compared to experimental value from \citet{Belov_gs}, Einstein $A$ coefficient (in s$^{-1}$), and sensitivity coefficient $T$ for vibrational ground state transitions of PH$_3$.}
%\resizebox{\linewidth}{!}{
\begin{tabular}{crrcrrrrrr}
\hline\hline\\[-3mm]
$\Gamma\p$ & \multicolumn{1}{c}{$J\p$} & \multicolumn{1}{c}{$K\p$}& \multicolumn{1}{c}{$\Gamma\pp$} & \multicolumn{1}{c}{$J\pp$}&\multicolumn{1}{c}{$K\pp$}&\multicolumn{1}{c}{$\nu_{\mathrm{calc}}$} & \multicolumn{1}{c}{$\Delta_{\mathrm{c}-\mathrm{e}}$} & \multicolumn{1}{c}{$A$} & \multicolumn{1}{c}{$T$} \\
	\hline\\[-2mm]
& &  &  &  & &\multicolumn{1}{l}{\textbf{Allowed}}  & & &  \\[1mm]
$A_2$ &  1& 0& $A_1$ & 0& 0&    266947.2&     2.7& 0.253E-04&  -0.99 \\
$E$ & 2& 1& $E$ & 1& 1&    533819.4&     4.2& 0.182E-03&  -0.99 \\
$A_1$ & 2& 0& $A_2$ & 1& 0&    533795.5&     0.9& 0.242E-03& -1.00 \\
$E$ &  3& 2& $E$ & 2& 2&    800586.8&     6.9& 0.486E-03& -1.00 \\
$E$ &  3& 1& $E$ & 2& 1&    800490.8&     3.7& 0.778E-03&  -0.99 \\
$A_2$ & 3& 0& $A_1$ & 2& 0&    800463.8&     7.7& 0.875E-03&  -0.99 \\
$A_1$ & 4& 3& $A_2$ & 3& 3&   1067210.2&     3.9& 0.940E-03&  -0.99 \\
$A_2$ & 4& 3& $A_1$ & 3& 3&   1067210.2&     3.9& 0.940E-03&  -0.99 \\
$E$ &  4& 2& $E$ & 3& 2&   1067006.3&     6.0& 0.161E-02&  -0.99 \\
$E$ &  4& 1& $E$ & 3& 1&   1066886.4&     9.5& 0.201E-02& -1.00 \\
$A_1$ &  4& 0& $A_2$ & 3& 0&   1066844.4&     8.5& 0.215E-02& -1.00 \\[1mm]
&  &  &  &   & & \multicolumn{1}{l}{\textbf{Forbidden}}  & & & \\[1mm]
$E$ &  6& 1& $E$ & 6& 2&     47409.2&    18.0& 0.780E-12&  -0.87 \\
$E$ &  7& 1& $E$ & 7& 2&     47199.3&    20.7& 0.140E-11&  -0.95 \\
$E$ &  8& 1& $E$ & 8& 2&     46962.5&    23.4& 0.232E-11&  -0.96 \\
$E$ &  9& 1& $E$ & 9& 2&     46695.7&    24.2& 0.362E-11& -1.08 \\
$E$ & 10& 1& $E$ & 10& 2&     46404.9&    27.1& 0.540E-11&  -0.85 \\
$E$ & 11& 1& $E$ & 11& 2&     46090.1&    31.6& 0.775E-11&  -0.85 \\
$E$ & 12& 1& $E$ & 12& 2&     45748.3&    33.5& 0.108E-10&  -0.86 \\
$E$ & 13& 1& $E$ & 13& 2&     45382.6&    34.7& 0.146E-10&  -0.83 \\
$E$ & 14& 1& $E$ & 14& 2&     44995.8&    37.2& 0.193E-10&  -0.96 \\
$E$ & 15& 1& $E$ & 15& 2&     44591.1&    42.2& 0.251E-10&  -0.92 \\
$A_2$ &  3& 0& $A_1$ & 3& 3&    143750.5&    48.9& 0.152E-11&  -0.98 \\
$A_1$ &  4& 0& $A_2$ & 4& 3&    143384.7&    53.7& 0.636E-11& -1.01 \\
$A_2$ &  5& 0& $A_1$ & 5& 3&    142923.1&    53.1& 0.169E-10& -1.01 \\
$A_1$ &  6& 0& $A_2$ & 6& 3&    142377.4&    58.4& 0.361E-10&  -0.96 \\
$A_2$ &  7& 0& $A_1$ & 7& 3&    141744.9&    65.8& 0.674E-10&  -0.96 \\
$A_1$ &  8& 0& $A_2$ & 8& 3&    141022.4&    71.9& 0.115E-09&  -0.96 \\
$A_2$ &  9& 0& $A_1$ & 9& 3&    140209.9&    76.8& 0.182E-09&  -0.96 \\
$A_1$ & 10& 0& $A_2$ &10& 3&    139307.6&    81.0& 0.275E-09&  -0.97 \\
$A_2$ & 11& 0& $A_1$ & 11& 3&    138318.2&    90.2& 0.398E-09&  -0.95 \\
$A_1$ & 12& 0& $A_2$ & 12& 3&    137230.0&    95.3& 0.557E-09&  -0.93 \\
$A_2$ & 13& 0& $A_1$ &13& 3&    136045.8&   104.4& 0.756E-09&  -0.90 \\
$A_1$ & 14& 0& $A_2$ &14& 3&    134750.7&   109.3& 0.100E-08&  -0.93 \\
$E$ &  6& 2& $E$ & 6& 5&    333977.8&   137.8& 0.903E-10&  -0.99 \\
$E$ &  7& 2& $E$ & 7& 5&    332493.8&   148.9& 0.225E-09&  -0.97 \\
$E$ &  8& 2& $E$ & 8& 5&    330815.0&   163.3& 0.452E-09&  -0.97 \\
$E$ &  9& 2& $E$ & 9& 5&    328941.3&   176.9& 0.801E-09&  -0.96 \\
$E$ & 10& 2& $E$ &10& 5&    326884.7&   194.4& 0.130E-08&  -0.98 \\
$E$ & 11& 2& $E$ &11& 5&    324645.2&   209.3& 0.199E-08&  -0.95 \\
$E$ & 12& 2& $E$ &12& 5&    322237.9&   228.9& 0.290E-08&  -0.95 \\
$E$ & 13& 2& $E$ &13& 5&    319665.7&   247.9& 0.406E-08&  -0.96 \\
$E$ & 14& 2& $E$ &14& 5&    316940.6&   268.7& 0.552E-08&  -0.93 \\
$E$ & 15& 2& $E$ &15& 5&    314068.6&   288.1& 0.731E-08&  -0.92 \\
$A_1$ & 7& 3&$A_2$ &  7& 6&    429296.8&   188.4& 0.249E-09&  -0.99 \\
$A_2$ &  7& 3& $A_1$ & 7& 6&    429284.8&   189.2& 0.249E-09& -1.00 \\
$A_1$ & 8& 3& $A_2$ & 8& 6&    427132.3&   205.7& 0.613E-09&  -0.98 \\
$A_2$ &  8& 3& $A_1$ & 8& 6&    427105.3&   207.2& 0.613E-09&  -0.98 \\
$A_1$ & 9& 3&$A_2$ &  9& 6&    424728.0&   227.1& 0.122E-08&  -0.98 \\
$A_2$ &  9& 3& $A_1$ & 9& 6&    424671.0&   227.0& 0.122E-08&  -0.98 \\
$A_1$ &10& 3&$A_2$ & 10& 6&    422092.8&   249.8& 0.213E-08&  -0.96 \\
$A_2$ & 10& 3& $A_1$ &10& 6&    421984.9&   247.2& 0.213E-08&  -0.96 \\
$A_1$ &11& 3&$A_2$ & 11& 6&    419238.8&   271.7& 0.343E-08&  -0.96 \\
$A_2$ & 11& 3& $A_1$ &11& 6&    419052.9&   269.8& 0.343E-08&  -0.96 \\
$A_1$ &12& 3&$A_2$ & 12& 6&    416186.9&   297.6& 0.519E-08&  -0.95 \\
$A_2$ & 12& 3& $A_1$ &12& 6&    415878.1&   294.8& 0.518E-08&  -0.94 \\
$A_1$ &13& 3&$A_2$ & 13& 6&    412949.1&   320.2& 0.748E-08&  -0.94 \\
$A_2$ & 13& 3& $A_1$ &13& 6&    412457.5&   316.6& 0.747E-08&  -0.95 \\
$A_1$ &14& 3&$A_2$ & 14& 6&    409558.5&   350.2& 0.104E-07&  -0.94 \\
$A_2$ & 14& 3& $A_1$ &14& 6&    408797.0&   341.3& 0.104E-07&  -0.94 \\
$A_1$ &15& 3&$A_2$ & 15& 6&    406029.9&   377.3& 0.140E-07&  -0.93 \\
$A_2$ & 15& 3& $A_1$&15& 6&    404893.7&   366.7& 0.139E-07&  -0.92 \\
\hline\hline\\[-2mm]
\end{tabular}%}
\end{table}

 The most striking sensitivities are displayed by the $A_1\!-\!A_2$ doublets of PH$_3$. As is well known for a molecule with $\bm{C}_{3\mathrm{v}}\mathrm{(M)}$ symmetry, all rotation-vibration energy levels corresponding to the same $K\equiv|k|\neq\!0$ rotational quantum number and having overall $A_1,A_2$ symmetry are split into doublets due to different ro-vibrational interactions (see, for example, \citet{Chen}). For the nondegenerate vibrational states, the $A_1\!-\!A_2$ splittings occur for rotational levels with $K\!=\!3n$ $(n\!=\!1,2,\ldots)$. For the doubly degenerate fundamental vibrational states characterized by the vibrational angular momentum quantum number $\ell\ne 0$, the splittings occur for the $K\!=\!1,2,4,5,7,8\ldots$ levels.

 In Tables~\ref{tab:k=3} to \ref{tab:2nu4}, we have computed sensitivity coefficients for a large number of the $A_1\!-\!A_2$ doublets for low-lying vibrational states. The results suggest that sensitivities of the $A_1\!-\!A_2$ splittings for non-coinciding ro-vibrational states possess values dependent on the rotational quantum number $J$. For example, $T\approx -1.5,-2,-3$ for $k\!=\!1,2,3$, respectively (see Tables~\ref{tab:k=3}, \ref{tab:k=3a}, \ref{tab:k=1,2} and \ref{tab:k=1,2a}). It would be interesting to see if this trend is present in other molecules of $\bm{C}_{3\mathrm{v}}\mathrm{(M)}$ symmetry. For the sensitivities corresponding to coinciding states, there is a strong and irregular dependence on the $x\!-\!y$ Coriolis interaction that can produce values at least one order of magnitude larger than the respective Coriolis-free predictions. This behaviour is similar to that of NH$_3$~\citep{Spirko:2014,Owens:2015,Owens:NH3:PRA} and H$_3$O$^+$~\citep{15OwYuPo.H3Op}.

\begin{table}
\centering
\caption{\label{tab:k=3}Calculated and experimental $k\!=\!3$, $A_1\!-\!A_2$ splittings (in MHz) and their sensitivities in the ground (gs) and $\nu_2$ vibrational states of PH$_3$.}
\begin{tabular}{rrrrrrr}
\hline\hline\\[-3mm]
\multicolumn{1}{c}{$J$} & \multicolumn{1}{c}{$\nu_{\mathrm{exp}}$} & \multicolumn{1}{c}{$\nu_{\mathrm{calc}}$} & \multicolumn{1}{c}{$T$}& \multicolumn{1}{c}{$\nu_{\mathrm{exp}}$} & \multicolumn{1}{c}{$\nu_{\mathrm{calc}}$} & \multicolumn{1}{c}{$T$} \\
\hline\\[-2mm]
   &         &\multicolumn{1}{c}{\textbf{gs}}&  &      &    \multicolumn{1}{c}{${\bm{\nu_2}}$}& \\[1mm]
 4 & 0.43409$^a$ & 0.450 &      &  3.60$^b$ & 3.568 &     \\
 5 & 1.73413$^a$ & 1.769 &      & 13.096$^b$&13.371 &     \\
 6 & 5.19570$^a$ & 5.246 &      & 36.627$^b$&37.384 &     \\
 7 & 12.9690$^a$ &13.101 &      &  78$^c$   &86.640 & -3.6    \\
 8 & 28.4825$^a$ &28.780 & -2.9 & 174$^c$   &176.19 & -3.6    \\
 9 & 56.8550$^a$ &57.440 & -2.7 & 318$^c$   &325.12 & -3.5    \\
10 &         &106.46 & -2.90& 531$^c$   &557.13 & -3.4    \\
11 &         &185.90 & -3.00& 872$^c$   &900.40 & -3.3    \\
12 &         &309.03 & -3.02&1412$^c$   &1388.2 & -3.3    \\
13 &         &493.04 & -3.04&2009$^c$   &2059.6 & -3.3    \\
14 &         &759.76 & -3.03&2896$^c$   &2959.6 & -3.17   \\
15 &         &1136.1 & -3.02&4686$^c$   &4139.8 & -3.13   \\
16 &         &1654.8 & -3.03&           &5660.3 & -3.10   \\
17 &         &2355.8 & -3.03&           &7588.6 & -3.07   \\
18 &         &3285.9 & -3.03&           & 10002 & -3.06   \\
\hline\hline\\[-2mm]
\end{tabular}
\vspace*{1mm}
\\[-2mm]
{\footnotesize $^a$\citet{Davies}, $^b$\citet{Chen}, $^c$\citet{Papousek}.}
\end{table}

\begin{table}
\centering
\caption{\label{tab:k=3a}Calculated and experimental~\citep{Ulenikov} $k\!=\!3$, $A_1\!-\!A_2$ splittings (in MHz) and their sensitivities in the $\nu_1$ and $2\nu_4^{\ell=2}$ vibrational states of PH$_3$.}
\begin{tabular}{rrrrrrr}
\hline\hline\\[-3mm]
\multicolumn{1}{c}{$J$} & \multicolumn{1}{c}{$\nu_{\mathrm{exp}}$} & \multicolumn{1}{c}{$\nu_{\mathrm{calc}}$} & \multicolumn{1}{c}{$T$}& \multicolumn{1}{c}{$\nu_{\mathrm{exp}}$} & \multicolumn{1}{c}{$\nu_{\mathrm{calc}}$} & \multicolumn{1}{c}{$T$} \\
\hline\\[-2mm]
   &         &\multicolumn{1}{c}{${\bm{\nu_1}}$}&  &           &\multicolumn{1}{c}{${\bm{2\nu_{4}^{\ell=2}}}$}&        \\[1mm]
 4 &    & 2.10 &         &  573  & 503.6& -2.6  \\
 5 &    & 7.88 &         & 1811  & 1561 & -2.6  \\
 6 &    &21.62 &         & 3906  & 3337 & -2.5  \\
 7 &    &47.37 & -5      & 6763  & 5792 & -2.33 \\
 8 & 45 &87.18 &-3.8     & 9719  & 8885 & -2.25 \\
 9 &114 &140.0 &-3.84    & 14429 &12504 & -2.15 \\
10 &    &201.9 &-2.97    & 17649 &15241 & -1.54 \\
11 &195 &261.9 &-2.05    & 19460 &17519 & -3.17 \\
12 &342 &480.0 &-3.38    & 29539 &26118 & -2.15 \\
13 &255 &810.2 &-3.65    &       &32465 & -2.03 \\
14 &6706&959.2 & 21.3    &       &38781 & -1.98 \\
15 &    &6665  & 3.83    &       &45247 & -1.89 \\
16 &    &1756  &-14.17   &       &51606 & -1.86 \\
17 &    &3405  & -5.39   &       &57544 & -1.76 \\
18 &    &5880  &-10.61   &       &52187 &  0.15 \\
\hline\hline\\[-2mm]
\end{tabular}
\end{table}

\begin{table}
\centering
\caption{\label{tab:k=1,2}Calculated and experimental $k\!=\!1$ and $k\!=\!2$, $A_1\!-\!A_2$ splittings (in MHz) and their sensitivities in the $\nu_4$ vibrational state of PH$_3$.}
\begin{tabular}{rrrrrrr}
\hline\hline\\[-3mm]
\multicolumn{1}{c}{$J$} & \multicolumn{1}{c}{$\nu_{\mathrm{exp}}$} & \multicolumn{1}{c}{$\nu_{\mathrm{calc}}$} & \multicolumn{1}{c}{$T$}& \multicolumn{1}{c}{$\nu_{\mathrm{exp}}$} & \multicolumn{1}{c}{$\nu_{\mathrm{calc}}$} & \multicolumn{1}{c}{$T$} \\
\hline\\[-2mm]
   &         &\multicolumn{1}{c}{${\bf k\!=\!1}$}&  &           &\multicolumn{1}{c}{${\bf k\!=\!2}$}&        \\[1mm]
 1 & 10498.9$^a$& 10429.5& -1.51 &         &        &        \\
 2 & 31269.7$^a$& 31058.4& -1.50 &   30$^e$&   30.16&        \\
 3 & 61876.0$^b$& 61444.7& -1.49 &  150$^e$&  149.12& -2.51  \\
 4 &101889.7$^c$& 101152 & -1.48 &  438$^e$&  436.62& -2.37  \\
 5 & 150200$^d$ & 149480 & -1.47 &  986$^e$&  975.97& -2.35  \\
 6 & 207700$^d$ & 205747 & -1.45 & 1850$^e$&  1827.4& -2.37  \\
 7 & 271300$^d$ & 269219 & -1.44 & 3034$^e$&  3004.9& -2.32  \\
 8 & 342600$^d$ & 339148 & -1.43 & 4521$^e$&  4471.7& -2.22  \\
 9 & 419000$^d$ & 414689 & -1.40 & 6175$^e$&  6156.9& -2.17  \\
10 & 502300$^d$ & 497112 & -1.38 & 7989$^e$&  7980.1& -2.11  \\
11 & 588700$^d$ & 581763 & -1.37 & 9875$^e$&  9868.3& -2.03  \\
12 & 679300$^d$ & 671136 & -1.35 &11700$^d$&  11762 & -1.97  \\
13 & 772300$^d$ & 764108 & -1.34 &12600$^d$&  13613 & -1.93  \\
14 & 869600$^d$ & 860215 & -1.32 &17900$^d$&  15378 & -1.88  \\
15 & 967700$^d$ & 959068 & -1.31 &19600$^d$&  16754 & -1.68  \\
16 &            &1060200 & -1.30 &         &  12454 & -4.48  \\
17 &            &1162100 & -1.27 &         &  18172 & -1.81  \\
18 &            &1273400 & -1.27 &         &  19568 & -1.67  \\
\hline\hline\\[-2mm]
\end{tabular}
\vspace*{1mm}
\\[-2mm]
{\footnotesize $^a$\citet{Scappini}, $^b$\citet{Guarnieri}, $^c$\citet{Belov}, $^d$\citet{Tarrago}, $^e$\citet{Papousek}.}
\end{table}

\begin{table}
\centering
\caption{\label{tab:k=1,2a}Calculated and experimental~\citep{Ulenikov} $k\!=\!1$ and $k\!=\!2$, $A_1\!-\!A_2$ splittings (in MHz) and their sensitivities in the $\nu_3$ vibrational state of PH$_3$.}
\begin{tabular}{rrrrrrr}
\hline\hline\\[-3mm]
\multicolumn{1}{c}{$J$} & \multicolumn{1}{c}{$\nu_{\mathrm{exp}}$} & \multicolumn{1}{c}{$\nu_{\mathrm{calc}}$} & \multicolumn{1}{c}{$T$}& \multicolumn{1}{c}{$\nu_{\mathrm{exp}}$} & \multicolumn{1}{c}{$\nu_{\mathrm{calc}}$} & \multicolumn{1}{c}{$T$} \\
\hline\\[-2mm]
  &         &\multicolumn{1}{c}{${\bf k\!=\!1}$}&  &           &\multicolumn{1}{c}{${\bf k\!=\!2}$}&        \\[1mm]
 1 &  333&   533 & -1.57&     &  &  \\
 2 & 1004&  1596 & -1.45&   54&  59.48& -3.7     \\
 3 & 2015&  3177 & -1.46&  288& 293.89& -3.4    \\
 4 & 3385&  5256 & -1.45&  914& 864.93& -3.22    \\
 5 & 5081&  7788 & -1.46& 2078& 1959.3& -3.13    \\
 6 & 7069& 10686 & -1.44& 3990& 3747.7& -3.08    \\
 7 & 9156& 13763 & -1.38& 6799& 6301.6& -2.96    \\
 8 &11368& 16604 & -1.20&10460& 9453.3& -2.75    \\
 9 &12825& 18371 & -0.79&14558& 12583 & -2.36    \\
10 &12894& 37630 & -0.72&18386& 34468 & -1.61    \\
11 &10736& 41212 & -0.86&21441& 42221 & -1.95    \\
12 &39033& 45466 & -0.84&56796& 52692 & -2.17    \\
13 &44078& 49330 & -0.61&71518& 66005 & -2.41    \\
14 &49924& 51045 &  0.08&     & 83884 & -2.99    \\
15 &51214& 48059 &  0.58&     &224180 &  1.45    \\
16 &     & 34359 & -32.8&     &73345  & -7.15    \\
17 &     &128720 &  3.81&     &86681  &  5.34    \\
18 &     &128190 & -1.23&     &110930 & -3.44    \\
\hline\hline\\[-2mm]
\end{tabular}
\end{table}

\begin{table}
\centering
\caption{\label{tab:k=4,5a}Calculated and experimental~\citep{Ulenikov} $k\!=\!4$ and $k\!=\!5$, $A_1\!-\!A_2$ splittings (in MHz) and their sensitivities in the $\nu_3$ vibrational state of PH$_3$.}
\begin{tabular}{rrrrrrr}
\hline\hline\\[-3mm]
\multicolumn{1}{c}{$J$} & \multicolumn{1}{c}{$\nu_{\mathrm{exp}}$} & \multicolumn{1}{c}{$\nu_{\mathrm{calc}}$} & \multicolumn{1}{c}{$T$}& \multicolumn{1}{c}{$\nu_{\mathrm{exp}}$} & \multicolumn{1}{c}{$\nu_{\mathrm{calc}}$} & \multicolumn{1}{c}{$T$} \\
\hline\\[-2mm]
   &         &\multicolumn{1}{c}{${\bf k\!=\!4}$}&  &           &\multicolumn{1}{c}{${\bf k\!=\!5}$}&        \\[1mm]
 8 &      &  97.79&-2.94&     &        &      \\
 9 &      &  295.8&-2.70&     &   &   \\
10 &  563 &  889.3&-2.81&     &  135.1 & -2.87\\
11 & 2215 &  2988 &-2.36& 186 &  267.1 & -2.81\\
12 & 5546 & 35905 &-2.45& 216 &  278.9 &  2.02\\
13 & 3439 &  6242 &-2.91& 306 &   16.1 & 461\\
14 & 3663 &  6172 &-5.8 &     &   6171 &-25.50\\
15 &      &  6098 &-5.4 &     &  15175 & 18.1\\
\hline\hline\\[-2mm]
\end{tabular}
\end{table}

\begin{table}
\centering
\caption{\label{tab:k=4}Calculated and experimental $k\!=\!4$ and $k\!=\!7$, $A_1\!-\!A_2$ splittings (in MHz) and their sensitivities in the $\nu_4$ vibrational state of PH$_3$.}
\begin{tabular}{rrrrrrr}
\hline\hline\\[-3mm]
\multicolumn{1}{c}{$J$} & \multicolumn{1}{c}{$\nu_{\mathrm{exp}}$} & \multicolumn{1}{c}{$\nu_{\mathrm{calc}}$} & \multicolumn{1}{c}{$T$}& \multicolumn{1}{c}{$\nu_{\mathrm{exp}}$} & \multicolumn{1}{c}{$\nu_{\mathrm{calc}}$} & \multicolumn{1}{c}{$T$} \\
\hline\\[-2mm]
   &         &\multicolumn{1}{c}{${\bf k\!=\!4}$}&  &           &\multicolumn{1}{c}{${\bf k\!=\!7}$}&        \\[1mm]
 5 &  582$^a$& 582.08& -2.08 &        &      &        \\
 6 & 1292$^a$& 1310.2& -2.18 &        &      &        \\
 7 & 2278$^a$& 2418.6& -2.13 &        &      &        \\
 8 & 3897$^a$& 3916.4& -2.15 &        &  &     \\
 9 & 5762$^a$& 5788.4& -2.11 & 210$^c$&211.7 &  -5.25 \\
10 & 7971$^a$& 8019.5& -2.08 &1190$^c$&1363.3&   1.04 \\
11 &10530$^a$& 10610 & -2.07 & 618$^c$& 646.3&  -2.09 \\
12 &13730$^a$& 13580 & -2.06 & 651$^c$& 671.4&  -2.33 \\
13 &16793$^a$& 16966 & -2.06 & 767$^c$& 796.7&  -2.54 \\
14 &20686$^a$& 20821 & -2.08 & 959$^c$& 987.3&  -2.67 \\
15 &23800$^b$& 25211 & -2.09 &1157$^c$&1246.5&  -2.71 \\
16 &         & 30209 & -2.10 &        &1589.1&  -2.84 \\
17 &         & 35900 & -2.12 &        &2038.3&  -2.87 \\
18 &         & 42378 & -2.15 &        &2625.3&  -2.94 \\
\hline\hline\\[-2mm]
\end{tabular}
\vspace*{1mm}
\\[-2mm]
{\footnotesize $^a$\citet{Davies}, $^b$\citet{Chen}, $^c$\citet{Papousek}.}
\end{table}

\begin{table}
\centering
\caption{\label{tab:k=7,8a}Calculated and experimental~\citep{Ulenikov} $k\!=\!7$ and $k\!=\!8$, $A_1\!-\!A_2$ splittings (in MHz) and their sensitivities in the $\nu_3$ vibrational state of PH$_3$.}
\begin{tabular}{rrrrrrr}
\hline\hline\\[-3mm]
\multicolumn{1}{c}{$J$} & \multicolumn{1}{c}{$\nu_{\mathrm{exp}}$} & \multicolumn{1}{c}{$\nu_{\mathrm{calc}}$} & \multicolumn{1}{c}{$T$}& \multicolumn{1}{c}{$\nu_{\mathrm{exp}}$} & \multicolumn{1}{c}{$\nu_{\mathrm{calc}}$} & \multicolumn{1}{c}{$T$} \\
\hline\\[-2mm]
   &         &\multicolumn{1}{c}{${\bf k\!=\!7}$}&  &           &\multicolumn{1}{c}{${\bf k\!=\!8}$}&        \\[1mm]
12 &    &30.55  & -64 &1340 &743.7  & -91 \\
13 &    &1260170& 1.44&     &18210  &  34 \\
14 &1817& 8766  & 0.21&3136 &10565  &-4.91\\
15 &    &2585.5 & 27.6&     &1344.4 &-22.3\\
16 &    & 807.4 &-1.11&     &1370.2 & 11.7\\
17 &    &1485.1 &-22.9&     &5056.4 &-16.8\\
18 &    &1315.0 &-417 &     & 73.0  & -3.1\\
\hline\hline\\[-2mm]
\end{tabular}
\end{table}

\begin{table}
\centering
\caption{\label{tab:2nu4} Calculated and experimental~\citep{Ulenikov} $k\!=\!1$, $A_1\!-\!A_2$ splittings (in MHz) and their sensitivities in the $2\nu_4^{\ell=2}$ vibrational state of PH$_3$. The splitting $\nu=\Delta E_{A_1/A_2}=(E_{A_2}-E_{A_1}\cdot (-1)^J )$. The sensitivity $T_{\mathrm{exp}}$ is obtained using the frequencies from \citet{Ulenikov} instead of the computed values.}
\begin{tabular}{rrrrr}
\hline\hline\\[-3mm]
\multicolumn{1}{c}{$J$} & \multicolumn{1}{c}{$\nu_{\mathrm{calc}}$} &
\multicolumn{1}{c}{$T_{\mathrm{calc}}$} & \multicolumn{1}{c}{$\nu_{\mathrm{exp}}$} &
\multicolumn{1}{c}{$T_{\mathrm{exp}}$} \\
\hline\\[-2mm]
       1&$        221.7055  $&$      -2.5  $&$    216.7499  $&$       -2.6  $\\
       2&$        592.9445  $&$      -2.3  $&$    573.2032  $&$       -2.4  $\\
       3&$        653.0709  $&$      -1.8  $&$    604.0818  $&$       -1.9  $\\
       4&$        286.2628  $&$      -0.6  $&$    220.3475  $&$       -0.7  $\\
       5&$         48.8692  $&$       1.7  $&$    -62.0570  $&$        1.4  $\\
       6&$         31.0615  $&$      -4.3  $&$   -115.4201  $&$       -1.2  $\\
       7&$        150.8825  $&$      -3.8  $&$    -23.3838  $&$      -24.4  $\\
       8&$        370.1448  $&$      -3.1  $&$    160.3890  $&$       -7.2  $\\
       9&$        656.5515  $&$      -2.7  $&$    413.1140  $&$       -4.2  $\\
      10&$        964.8880  $&$      -2.3  $&$    757.2757  $&$       -3.0  $\\
      11&$       1240.5502  $&$      -2.0  $&$   1072.9572  $&$       -2.3  $\\
      12&$       1432.5733  $&$      -1.5  $&$   1087.6470  $&$       -1.9  $\\
      13&$       1495.6166  $&$      -3.1  $&$   1141.3099  $&$       -4.1  $\\
\hline\hline\\[-2mm]
\end{tabular}
\end{table}

\begin{figure}
\centering
\includegraphics{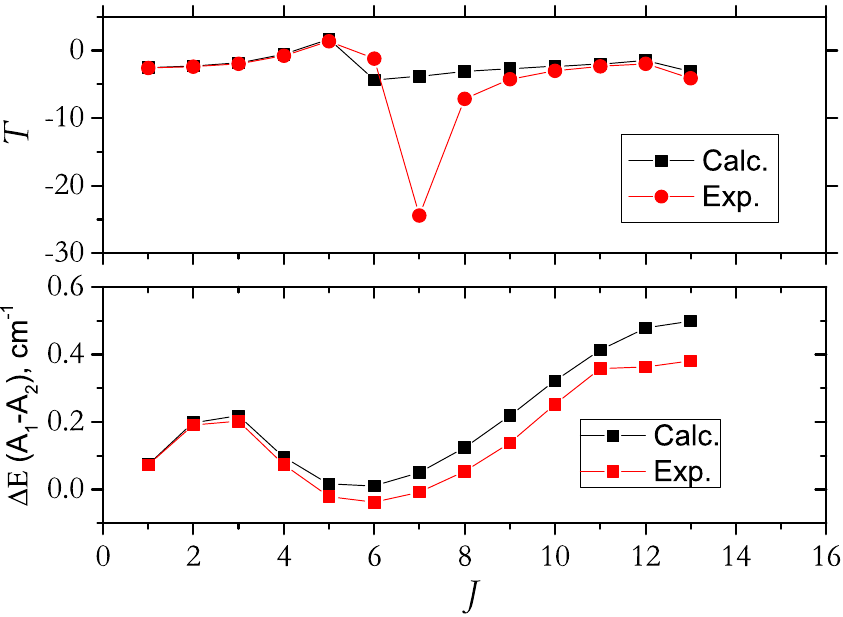}
\caption{\label{fig:2nu4} The $A_1$--$A_2$ splittings in the $2\nu_4^{\ell=2}$ state of PH$_3$ (lower panel) and the corresponding sensitivities $T$ (upper panel). The experimentally determined  energies by \citet{Ulenikov} were used in Eq.~\eqref{eq.T} to estimate the $T_{\mathrm{exp}}$ values. }
\end{figure}

A detailed study of the $A_1\!-\!A_2$ splittings in the $2\nu_4^{\ell=2}$ state was presented by \citet{Ulenikov} where it was shown that the dependence of the splitting on $J$ in the $K=1$ rotational sub-levels was anomalous between $J=3$--$8$. This anomaly is caused by an interaction with the closely lying $2\nu_4^{\ell=0}$ state ($K=0$). In Fig.~\ref{fig:2nu4} and Table~\ref{tab:2nu4} we show the $A_1\!-\!A_2$ splittings in the $2\nu_4^{\ell=2}$ state and corresponding sensitivity coefficients with respect to $J$. Aside from the $J=7$ sensitivity coefficient, which greatly increases when using the experimental frequency value, there is good agreement with the work of \citet{Ulenikov} and the sensitivities are highly anomalous.

It should be stated that for very energetically close coinciding states our variational approach may not be capable of a truly quantitative description. This is the reason why sensitivities have not been computed for certain extremely small $A_1\!-\!A_2$ splittings. Also, where computed frequencies noticeably differ from the experimental values the resultant sensitivities should only be regarded as illustrative, for example, in Table~\ref{tab:k=7,8a}. We have encountered this problem before~\citep{Owens:NH3:PRA} and whilst the underlying numerical derivatives are relatively stable, it is safer to regard the predicted sensitivity coefficients with caution.  Despite this, a large number of the computed $A_1\!-\!A_2$ splittings are in good agreement with experiment and, more importantly, reside in the radio frequency region.

\section{Conclusion}
\label{sec:conc}

 The sensitivity of the rotation-vibration spectrum of PH$_3$ to a possible variation of $\mu$ has been probed using an accurate variational approach. Calculations utilized the nuclear motion program \textsc{trove} in conjunction with an established empirically refined PES and \textit{ab initio} DMS. The low-lying vibrational states were studied as these play an important role in phosphine excitation in the carbon star envelope IRC +10216. Whilst the majority of computed sensitivity coefficients assumed their expected values, anomalous sensitivities were displayed by the $A_1\!-\!A_2$ splittings in the $\nu_2/\nu_4$, $\nu_1/\nu_3$ and $2\nu_4^{\ell=0}/2\nu_4^{\ell=2}$ manifolds. This behaviour arises due to strong Coriolis interactions between states and may be present in other molecules with $\bm{C}_{3\mathrm{v}}\mathrm{(M)}$ symmetry. The fact that molecules with highly sensitive transitions such as ammonia are already being used in advanced terrestrial experiments~\citep{Bethlem:2016} suggests that PH$_3$ may not be a primary candidate for constraining $\mu$ in laboratory studies. Its merit as a probe for a drifting constant is more likely to be in cosmological settings as it is a relevant astrophysical molecule with a well documented spectrum and a negligible hyperfine splitting~\citep{Mueller}. However, it is hard to comment on the necessary conditions for its detection since its presence and formation are not well understood (see the discussion by \citet{15SoAlTe.PH3} and references therein). Despite this, PH$_3$ as a model system shows that the splittings caused by higher-order rotation-vibration interactions, which are essentially low-frequency transitions that can be measured using radio telescopes, have real potential for investigating a possible variation of $\mu$.

\section*{Acknowledgments}

A.O. acknowledges support from the Deutsche Forschungsgemeinschaft (DFG) through the excellence cluster ``The Hamburg Center for Ultrafast Imaging – Structure, Dynamics and Control of Matter at the
Atomic Scale'' (CUI, EXC1074). S.Y. acknowledges support from the COST action MOLIM No. CM1405. V.S. acknowledges the research project RVO:61388963 (IOCB) and support from the Czech Science Foundation (grant P209/15-10267S).

\bibliographystyle{mn2e}
%\bibliography{ph3_mnras_a1a2}

\label{lastpage}

\end{document}